\documentclass[twoside]{ilcws07}
\usepackage[latin1]{inputenc}
\usepackage[dvips]{graphicx,epsfig,color}
\usepackage{wrapfig,rotating}
\usepackage{amssymb,amsmath,array}
\usepackage{cite} 
\def\lsim{\raise0.3ex\hbox{$\;<$\kern-0.75em\raise-1.1ex\hbox{$\sim\;$}}}
\def\gsim{\raise0.3ex\hbox{$\;>$\kern-0.75em\raise-1.1ex\hbox{$\sim\;$}}}
\begin{document}
\title{
%%%%   Paper title goes here  %%%%%%%%%%%%%%
How light can the lightest neutralino be?
} %% 
%***********************************************************************
% AUTHORS INFORMATION AREA
%***********************************************************************
\author{
H.~K.~Dreiner$^{1}$, S.~Heinemeyer$^{2}$, O.~Kittel\thanks{Speaker}~$^{1}$,
U.~Langenfeld$^{1}$, A.~M.~Weber$^{3}$, and G.~Weiglein$^{4}$
% DO NOT MODIFY THE FOLLOWING '\vspace' ARGUMENT
\vspace{.3cm}\\
1- Physikalisches Institut der Universit\"at Bonn \\
Nu\ss allee 12, 53115 Bonn - Germany
\vspace{.1cm}\\
2- Instituto de Fisica de Cantabria (CSIC-UC), \\
Santander - Spain
\vspace{.1cm}\\
3- Max-Planck-Institut f\"ur Physik (Werner-Heisenberg-Institut) \\
F\"ohringer Ring 6, 80805 M\"unchen - Germany
\vspace{.1cm}\\
4- University of Durham - IPPP\\
Durham DH1~3LE - UK
}
%%***********************************************************************

% END OF AUTHORS INFORMATION AREA

%***********************************************************************

\maketitle

\begin{abstract}
We show that 
in the Minimal Supersymmetric Standard Model,
the mass of the lightest neutralino is experimentally unconstrained
if the GUT relation between the gaugino mass parameters 
$M_1$ and $M_2$ is dropped.
We discuss what the impact of light or massless
neutralinos would be on their production at LEP, as well
as on electroweak precision data and rare decays.
\begin{picture}(10,10)
\put(40,260){IPPP/07/27, DCPT/07/54, BONN-TH-2007-03}
\end{picture}
\end{abstract}

\section{Introduction}

In the Minimal
Supersymmetric Standard Model (MSSM)~\cite{Haber:1984rc},
the masses and mixings of the neutralinos and charginos
are given by their mass matrices~\cite{Haber:1984rc,Yao:2006px}
%predicts Supersymmetric (SUSY) partners to the SM particles 
%in the range of a few hundred GeV.
%the four neutralinos $\tilde\chi^0_i$ are the fermionic 
%Supersymmetric (SUSY) partners of the neutral gauge
%and CP-even Higgs bosons.
%The two charginos $\tilde\chi^\pm_j$
%are the SUSY partners of the charged gauge and Higgs bosons.
%At tree level, the neutralino and chargino sector 
%only depend on the 
%$U(1)_Y$ and $SU(2)_L$ gaugino masses $M_1$ and $M_2$, respectively,
%the higgsino mass parameter $\mu$,
%and the ratio $\tan\beta=v_2/v_1$ of the vacuum expectation values 
%of the two Higgs fields.
%The masses and mixings of the neutralinos and charginos
%are given by their mixing matrices~\cite{Haber:1984rc,Yao:2006px}
\begin{equation}
%\tiny
{\mathcal M}_0
= M_Z\!\left(\!\!\begin {array}{cccc} M_1/M_Z&0&
          -s_\theta c_\beta &
\phantom{-}s_\theta s_\beta  \\
0&M_2/M_Z& 
\phantom{-}c_\theta c_\beta  &
          -c_\theta s_\beta \\
          -s_\theta c_\beta  &
\phantom{-}c_\theta c_\beta  &0&-\mu/M_Z\\
\phantom{-}s_\theta s_\beta &
-c_\theta s_\beta &-\mu/M_Z&0
\end{array}
\!\!\right), \,
{\mathcal M}_\pm=M_W\!\left(\!\!
\begin{array}{cc}
M_2/M_W &\sqrt{2} s_\beta\\
\sqrt{2}c_\beta&\mu/M_W 
\end{array}
\!\!\right), 
\end{equation}
respectively,
with $c_\beta= \cos\beta$, $s_\beta= \sin\beta$ and
$c_\theta=\cos\theta_w$, $s_\theta=\sin\theta_w$,
with the weak mixing angle $\theta_w$.
Besides the masses  
of the $W$ and $Z$ boson, $M_W$ and $M_Z$, respectively,
the neutralino and chargino sectors 
at tree level only depend on the 
$U(1)_Y$ and $SU(2)_L$ gaugino masses $M_1$ and $M_2$, respectively,
the higgsino mass parameter $\mu$,
and the ratio $\tan\beta=v_2/v_1$ of the vacuum expectation values 
of the two Higgs fields.
The neutralino (chargino) masses are the 
square roots of the eigenvalues of
${\mathcal M}_0 { \mathcal M}_0^\dagger$ 
(${\mathcal M}_\pm {\mathcal M}_\pm^\dagger$)~\cite{Yao:2006px}.
%Colliders have searched for the SUSY particles, but only
%limits on their masses could be set.
The LEP limit on the chargino mass is 
$m_{\tilde\chi^\pm_1}\gsim100$~GeV~\cite{Yao:2006px},
from which follows that $M_2,|\mu| \gsim 100$~GeV.
If the GUT relation  $M_1 = 5/3 \tan^2(\theta_w)M_2\approx 0.5\, M_2$
is assumed, then $M_1\gsim 50$~GeV, such that the
lightest neutralino mass is constrained to
$m_{\tilde\chi^0_1}\gsim 50$~GeV~\cite{Yao:2006px}.
However, if one drops the GUT relation, $M_1$ is an independent
parameter, allowing to tune the neutralino mass 
determined from the lowest-order mass matrix ${\mathcal M}_0$
freely~\cite{Choudhury:1999tn,Gogoladze:2002xp,
cosmology,dreineretal,longfield}.
%%xfreely~\cite{Choudhury:1999tn,Gogoladze:2002xp,
%%xHooper:2002nq,Bottino,Belanger:2003wb,longfield}.
 The neutralino  mass is identically zero 
%$m_{\tilde\chi^0_1}=0$~GeV, 
for~\cite{Gogoladze:2002xp}
\begin{eqnarray}
{\rm det}({\mathcal M}_0)=0\quad \Rightarrow M_1 = 
\frac{M^2_Z M_2 \sin^2\theta_w \sin(2\beta)}
{\mu M_2 - M_Z^2  \cos^2\theta_w \sin(2\beta)}
%\approx \frac{m_Z^2}{\mu} \sw[2]\sin(2\beta) 
\approx 0.05 \frac{M_Z^2}{\mu} = \mathcal{O}(1\,\mathrm{GeV}). 
\label{Eq:zeromass}
\end{eqnarray}
%Similar relations can be obtained for complex $\mu$, $M_1$.
%%%%%%%%%%%%%%%%%%%%%%%%%%%%%%%%%%%%%%%%%%%%%%%%%%%%%%%%%%%%%%%%%%%%%%%%%%%%%%
\begin{figure}[t]
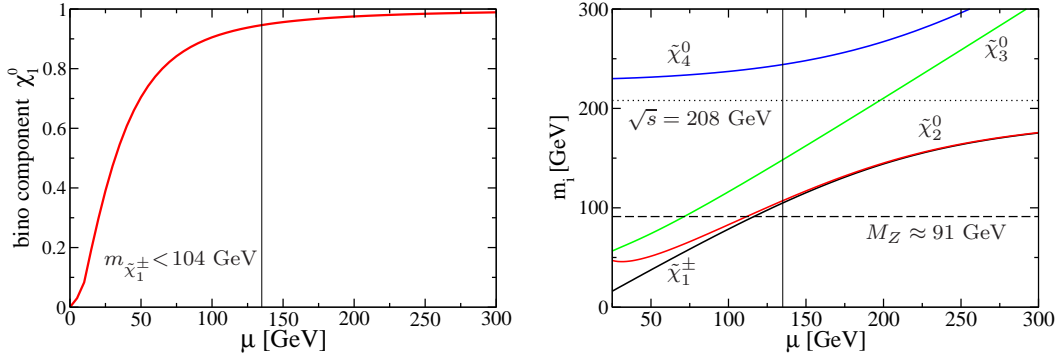

\begin{picture}(200,80)
\put(-5,-10){\includegraphics{./N11mu_200.eps}}
\put(200,-10){\includegraphics{./plot_masses.eps}}
\put(32,30){\footnotesize $m_{\tilde\chi_1^\pm}\!\!<\!104$~GeV}
\put(340,80){\footnotesize $\tilde\chi_2^0$}
\put(365,110){\footnotesize $\tilde\chi_3^0$}
\put(245,108){\footnotesize $\tilde\chi_4^0$}
\put(245,25){\footnotesize $\tilde\chi_1^\pm$}
\put(230,83){\footnotesize $\sqrt s= 208$~GeV}
\put(320,40){\footnotesize $M_Z\approx 91$~GeV}
%
%\put(200,-10){200}
%\put(250,-10){250}
%\put(300,-10){300}
%\put(350,-10){350}
%\put(400,-10){400}
%
%\put(400,0){0}
%\put(400,50){50}
%\put(400,100){100}
%
\end{picture}
\caption{\small 
        Bino admixture of $\tilde\chi_1^0$ (left plot) and masses of 
        charginos and neutralinos (right plot)
        for $M_2=200$~GeV, $\tan\beta=10$,
        and $M_1$ as given in Eq.~(\ref{Eq:zeromass}), such that
        $m_{\tilde\chi_1^0}=0$~GeV~\cite{dreineretal}.
        Left to the vertical lines at $\mu\approx135$~GeV, the chargino
        mass is $m_{\tilde\chi_1^\pm}< 104$~GeV.
        In the right panel, the dotted line indicates the reach of LEP2
        ($\sqrt s= 208$~GeV)
        for $e^+e^- \to \tilde\chi_1^0\tilde\chi_i^0$ production,
        and the dashed line indicates
        the mass of the Z boson,  $M_Z\approx 91$~GeV.
}
\label{fig:neutmixandmass}
\end{figure}
%%%%%%%%%%%%%%%%%%%%%%%%%%%%%%%%%%%%%%%%%%%%%%%%%%%%%%%%%%%%%%%%%%%%%%%%%%%
For $M_1\ll M_2,|\mu|$, the neutralino $\tilde\chi^0_1$ is mainly a bino,
i.e., it couples to hypercharge, and the 
masses of the other neutralinos and charginos are
of the order of $M_2$ and $|\mu|$,
see Fig.~\ref{fig:neutmixandmass}. 
In the following, we discuss bounds on the neutralino mass
from production at LEP and from precision 
observables~\cite{dreineretal, longfield}, 
as well as bounds from rare meson decays~\cite{daniel}. 
Finally, we summarize bounds from cosmology and
astrophysics~\cite{cosmology,dreineretal,longfield}.
%%Xastrophysics~\cite{Hooper:2002nq,Bottino,Belanger:2003wb,longfield}.

\section{Neutralino production at LEP}

\begin{figure}[t]
\begin{picture}(200,100)
\put(5,-10){\includegraphics{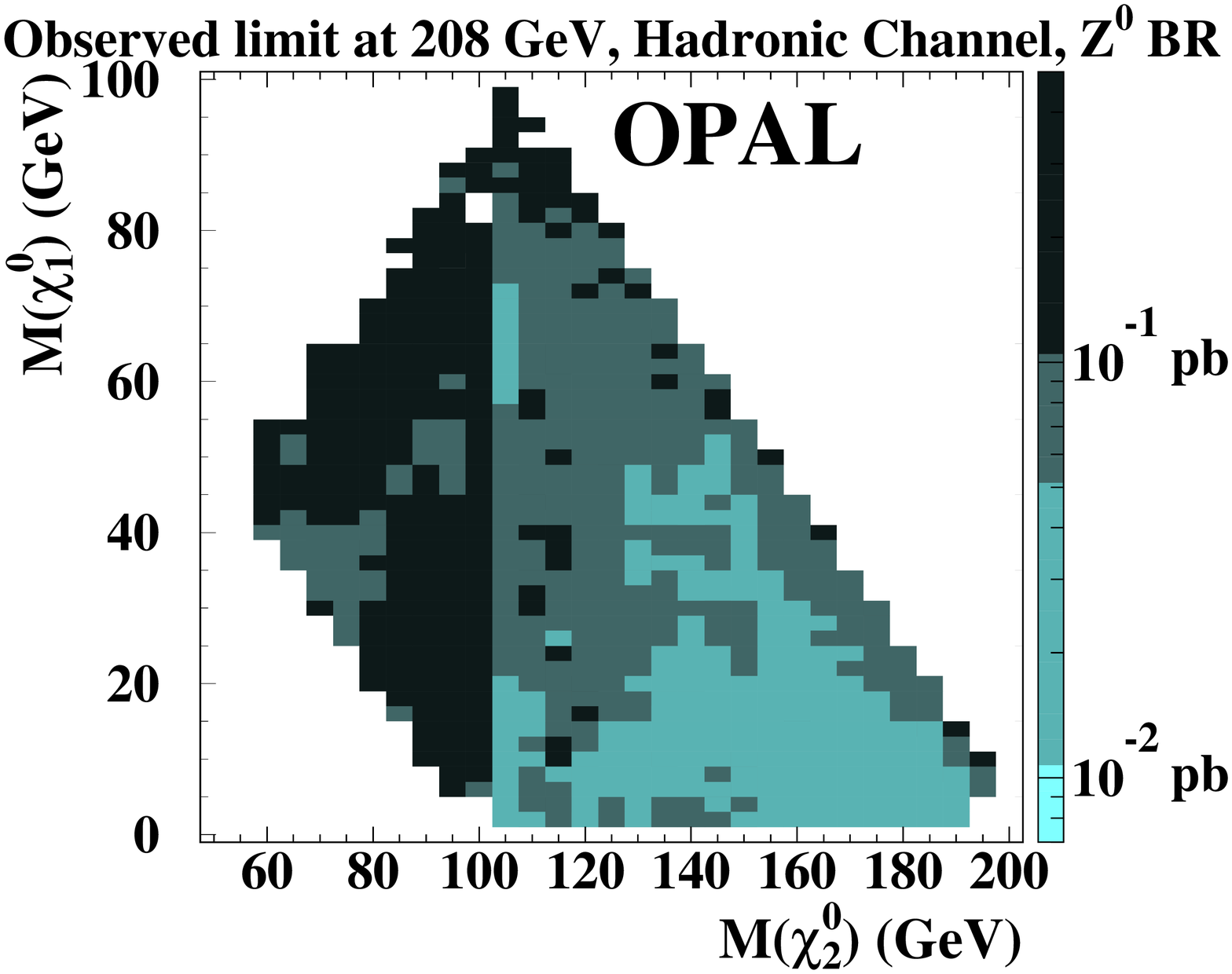}}
\put(200,-40){\includegraphics{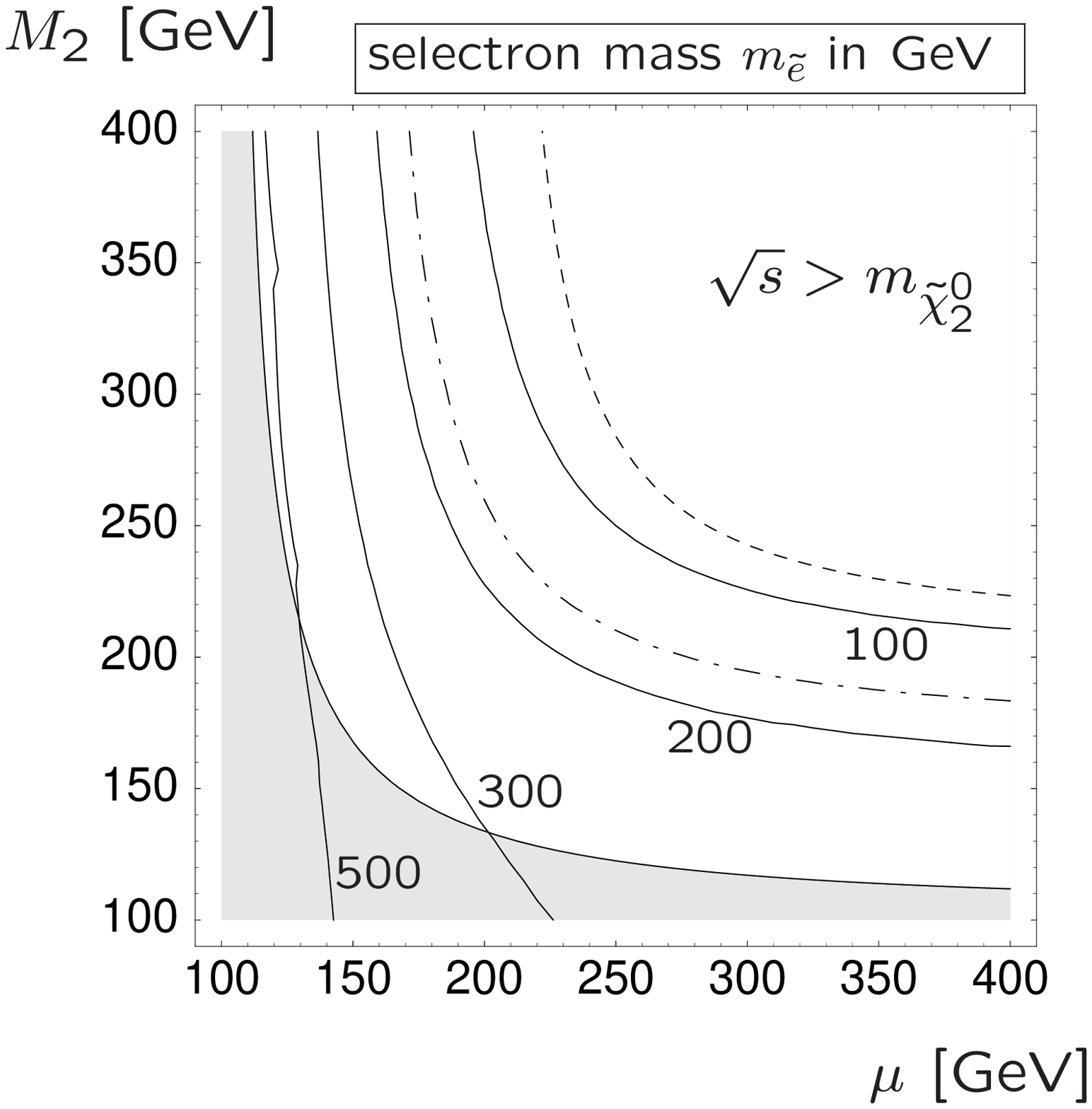}}
\put(40,-5){\small\bf (a)}
\put(260,-5){\small\bf (b)}
\end{picture}
\caption{\small 
        {\bf (a)} $95\%$ confidence limit on the cross section 
        $\sigma(e^+e^-\to\tilde\chi_1^0\tilde\chi_2^0)\times
        {\rm BR}(\tilde\chi_2^0\to Z\tilde\chi_1^0)\times
        {\rm BR}(Z\to q\bar q)$ with 
        ${\rm BR}(\tilde\chi_2^0\to Z\tilde\chi_1^0)=1$ at
        $\sqrt s = 208$~GeV,
        OPAL collaboration~\cite{Abbiendi:2003sc}.
        {\bf (b)} Contour lines in the $\mu$--$M_2$ plane 
        of the lower bounds on the selectron mass 
        $m_{\tilde e_R}=m_{\tilde e_L}=m_{\tilde e}$,
        such that $\sigma(e^+e^-\to\tilde\chi_1^0\tilde\chi_2^0)=70$~fb
        for $m_{\tilde\chi_1^0}=0$ with 
        $\tan\beta=10$~\cite{dreineretal}.
        The dashed line in {\bf (b)} is the kinematical limit
         $m_{\tilde\chi_2^0}= \sqrt s=208$~GeV,
        along the dot-dashed contour the relation
        $m_{\tilde e}=m_{\tilde\chi_2^0}$ holds.
}
\label{fig:OPALbounds}
\end{figure}

The OPAL collaboration~\cite{Abbiendi:2003sc} has
derived upper bounds on the topological neutralino production
cross section 
$\sigma(e^+e^-\to\tilde\chi_1^0\tilde\chi_2^0)\times
{\rm BR}(\tilde\chi_2^0\to Z\tilde\chi_1^0)\times
{\rm BR}(Z\to q\bar q)$ at LEP
with  $\sqrt s= 208$~GeV,
normalized such that 
${\rm BR}(\tilde\chi_2^0\to Z\tilde\chi_1^0)=1$.
Their observed limit at $95\%$ confidence level in the
$m_{\tilde\chi_1^0}$--$m_{\tilde\chi_2^0}$ plane is shown 
in Fig.~\ref{fig:OPALbounds}(a).
For $m_{\tilde\chi_1^0}=0$~GeV, one can roughly read off the upper limit
$\sigma(e^+e^-\to\tilde\chi_1^0\tilde\chi_1^0 q\bar q)<50$~fb,
or equivalently, 
since ${\rm BR}(Z\to q\bar q)\approx 70\%$,
$\sigma(e^+e^-\to\tilde\chi_1^0\tilde\chi_2^0)<70$~fb.
This is already a very tight bound, since typical 
neutralino production cross sections can be of the
order of $100$~fb. For bino-like neutralinos, the
main contribution to the cross section is due to
$\tilde e_R$ exchange.
Thus one can translate the OPAL bound on the neutralino production 
cross section into lower bounds on the selectron mass 
$m_{\tilde e_R}=m_{\tilde e_L}=m_{\tilde e}$, for
$m_{\tilde\chi_1^0}=0$.
In Fig.~\ref{fig:OPALbounds}(b) we show the contours
of $m_{\tilde e}$ in the $\mu$--$M_2$ plane,
such that along the contours 
$\sigma(e^+e^-\to\tilde\chi_1^0\tilde\chi_2^0)=70$~fb.
For example, 
for a fixed selectron mass of $m_{\tilde e}=300$~GeV, 
the area below the $300$~GeV contour in Fig.~\ref{fig:OPALbounds}(b) 
is excluded by LEP.
%\medskip

Another search channel at LEP is radiative neutralino production,
$e^+e^-\to\tilde\chi_1^0\tilde\chi_1^0\gamma$. 
Due to the 
large background from radiative neutrino production
$e^+e^-\to\nu\bar\nu\gamma$, we find that the significance is always
$S<0.1$ for $\mathcal L=100$~pb$^{-1}$
and $\sqrt s= 208$~GeV~\cite{Dreiner:2006sb,Dreiner:2007vm}.
At the ILC however, radiative neutralino production will be measurable,
due to the significant higher luminosity 
%$\mathcal L=500$~fb$^{-1}$
and the option of polarized 
beams~\cite{Dreiner:2006sb,Dreiner:2007vm,MoortgatPick:2005cw}.

%Since the OPAL analysis assumes heavy sleptons,
%the shown bounds are (conservative) upper bounds.
%The bounds could be tightened allowing for smaller sleptons masses.

%%%%%%%%%%%%%%%%%%%%%%%%%%%%%%%%%%%%%%%%%%%%%%%%%%%%%%%%%%%%%%%%%%%%%%%%%%%%%%
%%%%%%%%%%%%%%%%%%%%%%%%%%%%%%%%%%%%%%%%%%%%%%%%%%%%%%%%%%%%%%%%%%%%%%%%%%%

\section{Bounds from precision observables and rare decays}

\begin{figure}[t]
\begin{picture}(200,150)
\put(-30,0){\includegraphics{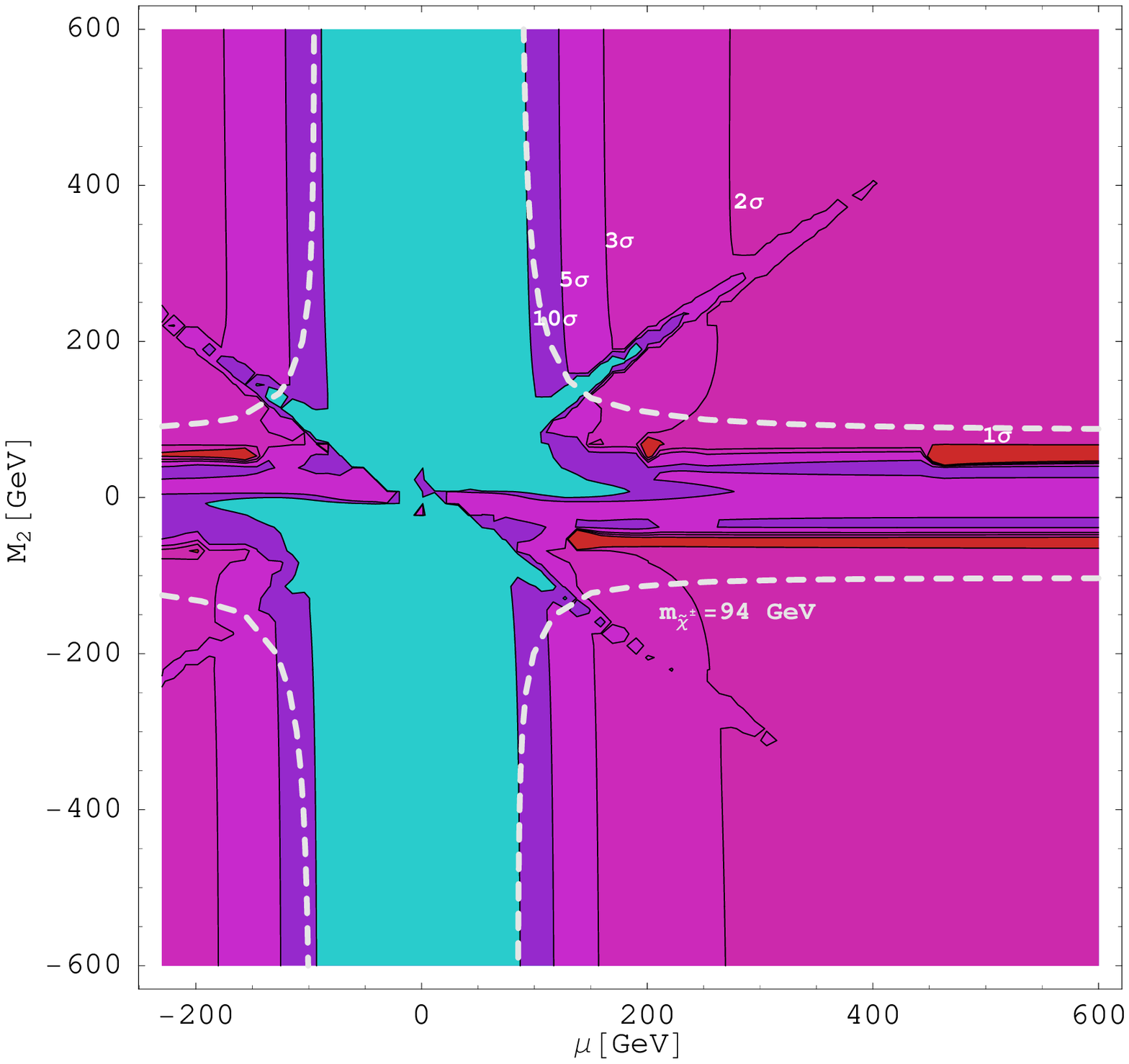}}
\put(170,0){\includegraphics{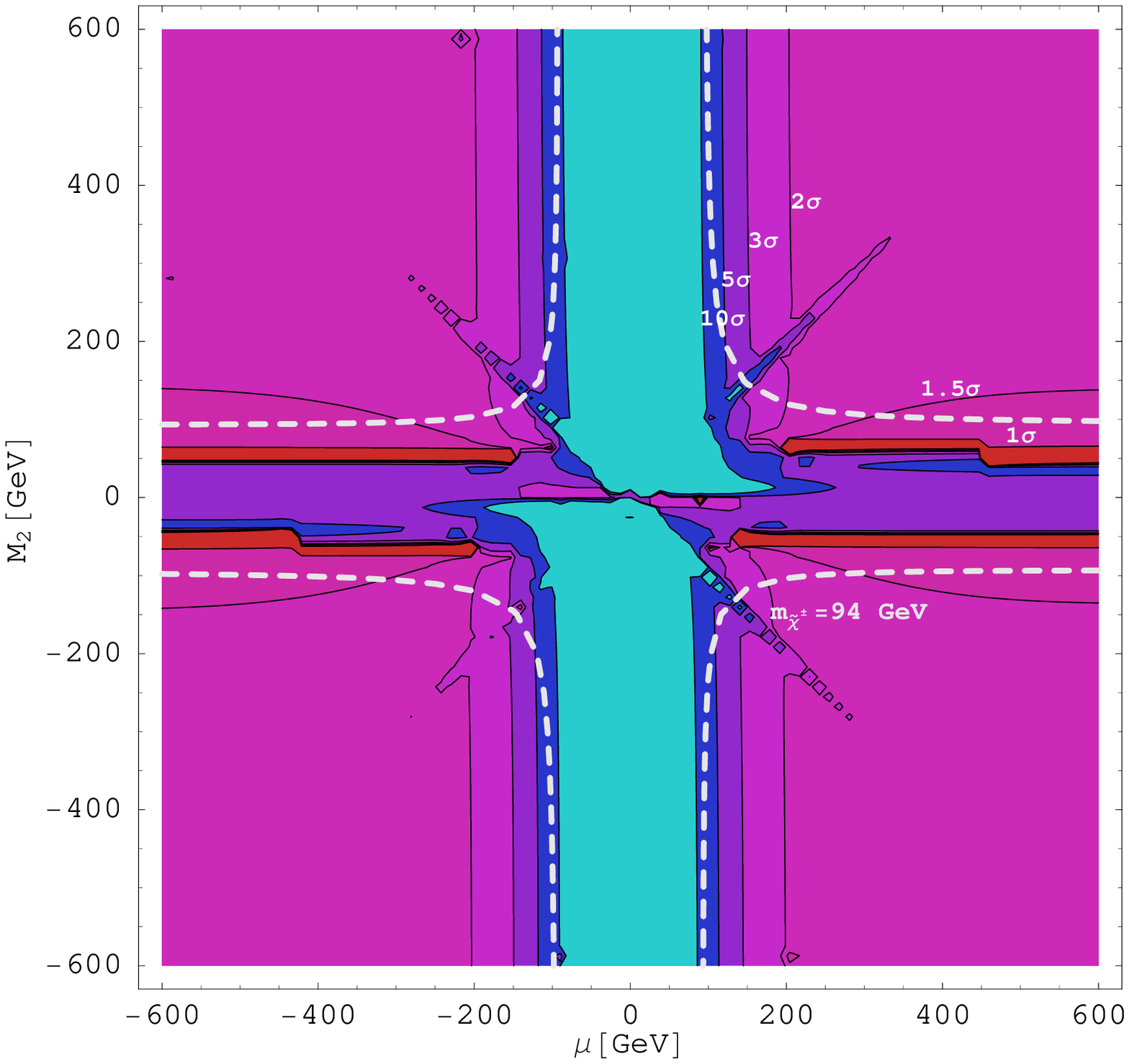}}
\put(30,-1){\small\bf (a)}
\put(240,-1){\small\bf (b)}
\end{picture}
\caption{\small 
         Contour lines in the $\mu$--$M_2$ plane for the difference 
         $\delta \Gamma =  
        (\Gamma_{\rm inv}-\Gamma_{\rm inv}^{\rm exp})/
        \Delta\Gamma$
         of theory prediction and
         experimental value of the invisible $Z$ width
         in units of the
         experimental error $\Delta\Gamma = 1.5~{\rm MeV}$,
         for  $m_{\tilde\chi_1^0} = 0$~GeV,
         $\tan\beta=10$, and {\bf (a)}
         $A_\tau=A_t=A_b=m_{\tilde g}=M_A=2M_{\tilde f}=500$~GeV,
          {\bf (b)}
          $A_\tau=A_t=A_b=m_{\tilde g}=M_A=M_{\tilde f}=600$~GeV.
          Along the dashed line 
          $m_{\tilde\chi_1^\pm}=94$~GeV.
}
\label{fig:invZwidth}
\end{figure}

The invisible $Z$ width $\Gamma_{\rm inv}$
is potentially very sensitive to a light or massless neutralino,
due to the contribution from $Z\to\tilde\chi_1^0\tilde\chi_1^0$.
However, a light neutralino is mainly bino-like for $|\mu|\gsim125$~GeV,
see Fig.~\ref{fig:neutmixandmass}. For a pure bino, the coupling to
the $Z$ boson vanishes at tree level.
In Fig.~\ref{fig:invZwidth}, we show the difference
$\delta \Gamma =  
        (\Gamma_{\rm inv}-\Gamma_{\rm inv}^{\rm exp})/
        \Delta\Gamma$
%$\delta \Gamma =  \Gamma_{\rm inv}-\Gamma_{\rm inv}^{\rm exp}$
from the measured invisible width
$\Gamma_{\rm inv}^{\rm exp} = 499.0\pm 
1.5$~MeV~\cite{Yao:2006px,lepewwg},
in units of the experimental
error $\Delta\Gamma=1.5$~MeV,
to the theoretical prediction $\Gamma_{\rm inv}$.
The calculations of $\Gamma_{\rm inv}$
include the full ${\mathcal O}(\alpha)$ SM and
MSSM contributions, supplemented with leading 
higher-order terms~\cite{ZObsMSSM}.
The deviation from the measured width
$\Gamma_{\rm inv}^{\rm exp}$ is larger than $5\sigma$
only for $|\mu|\lsim125$~GeV. For decreasing $|\mu|$, the increasing 
higgsino admixture leads to a 
non-negligible neutralino coupling to the $Z$ boson. 
Note that already the SM contribution to $\Gamma_{\rm inv}$
is more than $1\sigma$ larger than the
experimental value 
$\Gamma_{\rm inv}^{\rm exp}$~\cite{lepewwg,ZObsMSSM}.
%\begin{wrapfigure}{r}{0.5\columnwidth}
%\centerline{\includegraphics[width=0.45\columnwidth]{DeltaGammaZInv.eps}}
%\caption{
%}
%\label{fig:invZwidth}
%\end{wrapfigure}
%%%%%%%%%%%%%%%%%%%%%%%%%%%%%%%%%%%%%%%%%%%%%%%%%%%%%%%%%%%%%%%%%%%%%%%%%%%%%%
%%%%%%%%%%%%%%%%%%%%%%%%%%%%%%%%%%%%%%%%%%%%%%%%%%%%%%%%%%%%%%%%%%%%%%%%%%%

%\medskip
We have also studied the impact of a massless or light neutralino
on the $W$ boson mass, the effective leptonic weak mixing 
angle $\sin^2\theta_{\rm eff}$, the electric dipole moments of the electron,
neutron and mercury, and the anomalous magnetic moment of the muon
$(g-2)_\mu$, but have found no significant
constraints on the neutralino mass~\cite{dreineretal}.
Also  
rare decays like $b \to s\gamma$,
$\Upsilon(1S)\to \tilde\chi^0_1 \tilde\chi^0_1$~\cite{McElrath:2005bp},
$J/\Psi (B^0) \to \tilde\chi^0_1 \tilde\chi^0_1$,
$K[D,B]^+ \to \pi^+\tilde\chi^0_1 \tilde\chi^0_1$, do not constrain
$m_{\tilde\chi_1^0}$~\cite{daniel}.

\section{Bounds from cosmology and astrophysics}

The impact of a light neutralino on its thermal relic density
has previously been 
studied~\cite{cosmology,longfield}.
If the neutralino accounts for the dark matter, 
its mass has to be $m_{\tilde\chi_1^0} > 3\dots20$~GeV,
in order not to over-close the universe.
However, this bound can be evaded by allowing 
a small amount of R parity violation~\cite{Choudhury:1999tn}.
One would thus assume that the neutralinos 
are stable on the
time scale of collider experiments, but are not
stable on cosmological time scales.

Light neutralinos could be thermally produced inside a Supernova.
If their mean free path is of the order of the
Supernova core size or lager, the neutralinos escape
freely and lead to an additional cooling of the Supernova.
To be in agreement with
observations of 
the Kamiokande and IMB Collaborations
from SN 1987A, see Ref.~\cite{Dreiner:2003wh},
the cooling must not shorten the 
neutrino signal.
The energy that is emitted by the neutralinos
is much smaller than that emitted by the
neutrinos if 
$m_{\tilde\chi_1^0} \gsim 200$~MeV~\cite{Dreiner:2003wh},
with $m_{\tilde e}=500$~GeV.
For heavy sleptons,
$m_{\tilde e}\gsim1200$~GeV, however, 
no bound on  the neutralino mass 
can be set~\cite{longfield,Dreiner:2003wh}.

A very light neutralino would be a hot dark matter candidate.
The Cowsik-McClelland bound~\cite{Cowsik:1972gh} gives here
$m_{\tilde\chi_1^0} \lsim 1$~eV~\cite{longfield},
such that a light relativistic neutralino does not
disturb the formation of large structures in the universe.
Thus, a light or even massless neutralino
can be in agreement with constraints from
cosmology and astrophysics.

%\section{Summary and conclusions}
%
%In the MSSM, the neutralino can be chosen
%very light or even massless if no GUT relation between 
%$M_1$ and $M_2$ is assumed. 
%The light neutralino is then mainly a bino. 
%We have shown that there are no constraints on the neutralino
%mass from electroweak precision data and rare decays.
%The bounds on the cross section of neutralino production  
%at LEP $\sigma(e^+e^-\to\tilde\chi_1^0\tilde\chi_2^0)<70$~fb
%translate into bounds on the selectron mass
%$m_{\tilde e}<100\dots 500$~GeV.
%There are no bounds from 
%radiative neutralino production at LEP.
%We conclude that a light or even massless neutralino is allowed.

% ****************************************************************************
% BIBLIOGRAPHY AREA
% ****************************************************************************

\begin{footnotesize}
% IF YOU DO NOT USE BIBTEX, USE THE FOLLOWING SAMPLE SCHEME FOR THE REFERENCES
% ----------------------------------------------------------------------------

\end{footnotesize}

% ****************************************************************************
% END OF BIBLIOGRAPHY AREA
% ****************************************************************************

\end{document}